\documentclass[twocolumn]{revtex4-2}
\usepackage{amsmath}
\usepackage{graphicx}
\usepackage{textcomp}
\usepackage{gensymb}
\usepackage{subcaption}
\usepackage{dcolumn}

\renewcommand{\thesection}{\arabic{section}}
\renewcommand{\thesubsection}{\arabic{section}.\arabic{subsection}}

\begin{document}

\title{The Power Spectrum of Climate Change} 


\author{Albert Sneppen}
  \email{e-mail: a.sneppen@gmail.com (corresponding author)}
 \affiliation{Niels Bohr Institute, University of Copenhagen, Blegdamsvej 17, K\o benhavn \O~2200, Denmark.}
 \affiliation{Cosmic Dawn Center (DAWN)}

\begin{abstract}
Both global, intermediate and local scales of Climate Change have been studied extensively, but a unified diagnostic framework for examining all spatial scales concurrently has remained elusive. Here we present a new tool-set using spherical harmonics to examine climate change through surface temperature anomalies from 1850 to 2021 on spatial scales ranging from planetary to 50 km scales. We show that the observed temperature anomalies are accurately decomposed in spherical harmonics typically within 0.05 K.
This decomposition displays a remarkably simple dependence on spatial scale with a universal shape across seasons and decades. The decomposition separates two distinct regimes by a characteristic turnover-length of approximately 3000 km. The largest scales confirm established trends, while local fluctuations are consistent with 2-dimensional turbulence. 
We observe a downward cascade, from which it follows that climate change feeds increasing volatility on all spatial scales from $2.000$ to $50$ km. This increase is primarily driven by growing volatility along longitudes. \newline 
\end{abstract}

\flushbottom
\maketitle
\thispagestyle{empty}
\section{Introduction}

Climate Change affects the Earth on a global scale, but its impact exhibits inhomogeneities over many spatial scales. Both large scale features of climate change \cite{Screen2014,Kornhuber2019,temperature_data} and the evolution of local fluctuations in weather and climate \cite{Sutton2015,Vincze2017} have been extensively studied. However, the multi-scale nature of climate-change emphasises the need for a unified diagnostic examining correlations on all spatial scales \cite{Donges2009,Donges2015,Cecco2018,ZHANG2015}. Historically, Fourier decomposition of temperature-fluctuations has been utilised to determine the relative importance of different spatial scales \cite{Nastrom1985,Straus1999}. 
Such spectral domain analyses has revealed atmospheric fields (such as absolute temperature, wind velocity, pressure, etc.) are commonly scale-free over a wide dynamical range \cite{Lovejoy2008,Lovejoy2010,Chen2016,Cavanaugh2017}. However, the evolution of spectral coefficients remain unexplored across the time-scales of climate change. 


Therefore, this article introduces temperature anomalies to the spatial spectral domain and uniquely constrains the temporal evolution of this spectral regime. The analysis extends across a wide range of spatial scales from planetary to 50 km resolution with a monthly temporal resolution from 1850 to 2022. This is achieved by presenting the climate and its evolution on a range spatial scales through a collection of spherical harmonics. This basis is chosen across many fields of physics, because the visual nature of the different modes of spherical harmonics allows an interpretable decomposition of any scalar field, such as temperature fluctuations, on a sphere.

The significance of different spatial scales is described by the power spectrum of spherical harmonics. Therefore, we determine the spectrum's shape (see \S \ \ref{sec:power}), where large spatial scales confirm previous results within climatology and small scale features follow the analytical predictions of turbulence confined to two dimensions. Furthermore, we show that temperature anomalies on these different scales are causally connected because variations cascade from large to small spatial scales. 

We extend the analysis with a temporal evolution, where the contribution of each mode is detailed across time. Thus, in \S \ \ref{sec:lowest} we explore the evolution of the lowest modes with key results conforming to established observational trends within climatology. Furthermore, the time-evolution of the spectrum (see \S \ \ref{sec:time_evol}) quantifies how climate change results in an increased magnitude and frequency of local temperature fluctuations. This increase in volatility is not isotropic, but predominantly due to increased fluctuations along longitudes not latitudes (see \S \ \ref{sec:anisotropy}). 
Ultimately, spherical harmonics prove an intuitive framework in decomposing the local effects of global warming.

\begin{figure*}
    \centering
    \includegraphics[width=0.95\linewidth]{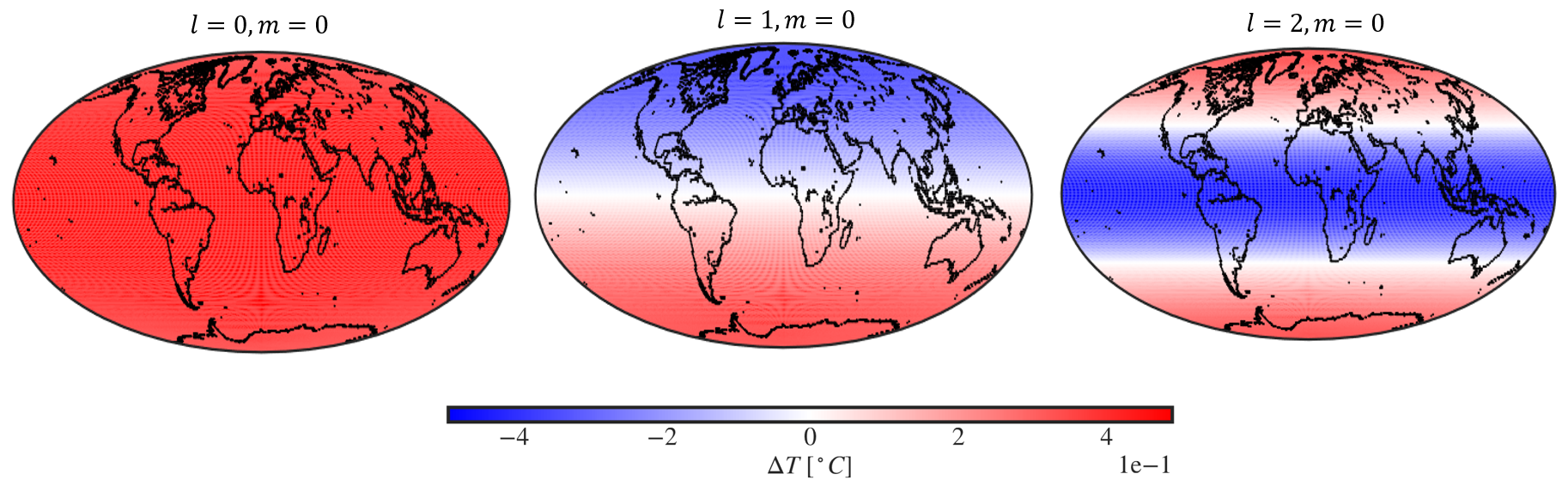}
    \caption{Projected visualisation of first three degrees ($l={0,1,2}$) of spherical harmonics with order $m=0$.
    \underline{Left:} The 0th spherical harmonics is an average value around the sphere. \underline{Center:} illustrates the different behaviour of the Northern and Southern hemisphere. \underline{Right:} emphasises the difference between the poles and the equator.}
    \label{fig:sph_vis}
\end{figure*}

\section{Methods}\label{sec:Theory}
\subsection{An Introduction to Spherical Harmonics}
Spherical harmonics are ortho-normal base functions on the surface of a sphere. They are in this analysis used to decompose the temperature anomalies across the planet, $T(\theta, \varphi,t)$ at any given time, $t$:
\begin{equation}\label{eq:decomp}
    T(\theta, \varphi,t) = \sum_{l=1}^\infty \sum_{m=-l}^l a_{l,m}(t) Y_l^m(\theta, \varphi)
\end{equation}
where $a_{l,m}$ is the coefficient of the spherical harmonic $Y_l^m(\theta, \varphi)$, which is given by the Legendre-polynomial $P_l^m(\cos\theta)$: 
\begin{equation}
    Y_l^m(\theta, \varphi) = (-1)^m \bigg[\frac{2l+1}{4\pi} \frac{(l-m)!}{(l+m)!} \bigg]^{1/2} P_l^m(\cos\theta) e^{im\varphi}
\end{equation}

The shape of $Y_l^m$ is described by the degree, $l$, which determines the number of oscillations around the sphere. In this context, the l'th harmonic is a description of fluctuations on angular scales $\theta \approx \frac{180}{l}$, which correspond to spatial scales of $\frac{20.000 km}{l}$. For instance, the monopole, $l=0$, gives the average value around the sphere; The dipole, $l=1$, shows the behaviour of the opposing hemispheres with different signs; The quadruple, $l=2$, describes the behaviour of the opposing hemispheres with the same sign (see Fig. \ref{fig:sph_vis}). 

Furthermore, the order, $m$ (ranging from $-l$ to $l$), determines whether the waves are oriented along longitudes ($m=0$) or latitudes ($m = \pm l$).  $a_{l,m}$ is the coefficient of the $Y_l^m$'th harmonic and are determined from the temperature field $T(\theta, \varphi,t)$:
\begin{equation}\label{eq:a_lim}
    a_{l,m}(t) = \int_{-1}^1\int_0^{2\pi} [Y_l^m(\theta, \phi)]^*\, T(\theta, \phi,t)\, d\phi \;  d(\cos \theta)
\end{equation}

Here, $|a_{l,m}|^2$ represents the contribution of any spherical harmonic to the decomposed function, so the significance of a degree is given by the power, $C_l$:
\begin{equation}\label{eq:pow}
    C_l(t) = \frac{1}{2l + 1}  \sum_{m=-l}^l|a_{l,m}(t)|^2 %
\end{equation}

Thus, investigating the power spectrum is a simple diagnostic tool for characterizing the magnitude of temperature fluctuations as a function of angular scale. 


%


\begin{figure*}
    \centering
    \includegraphics[width=\linewidth]{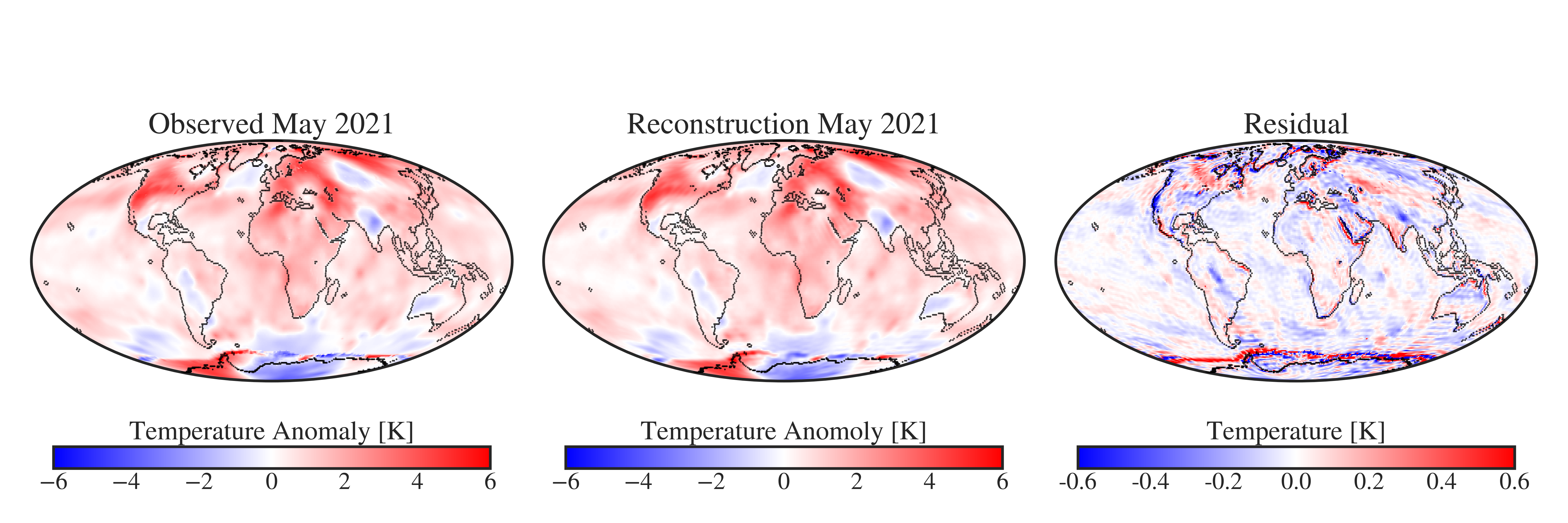}
    \caption{\underline{Left:} Observed temperature anomalies, \underline{Center:} Spherical Harmonical Reconstruction using all degrees $l<90$, \underline{Right:} Residual between observed and reconstructed for May 2021. Note, we emphasise discrepancies by increasing the contrast on residuals. The magnitude of residuals is typically small with 50\% of the surface being within 0.05 degrees.} 
    \label{fig:l_comb_mol_1}
\end{figure*}

\subsection{Data-sets and Potential Bias}
\label{sec: Part2}

The data was collected from The Berkeley Earth Land/Ocean Temperature Record \cite{temperature_data}. The data set is a combination of the Berkeley Earth land-surface temperature field \cite{land_temp}
and a reinterpolated version of the HadSST ocean temperature field \cite{Kennedy2019}. It provides the monthly temperature anomalies along with their uncertainties across the globe in a 360$\times$180 grid (longitude vs. latitude), spanning a period of 171 years between Jan 1850 and Dec 2021. The anomalies are reported relative to the locally inferred average temperature for that month over the period Jan 1951-Dec 1980 [for further details see \cite{temperature_data}]. 
Note, that both absolute temperature and temperature anomalies can be decomposed in spherical harmonics, but the former is affected by systematic shifts [for instance the dependence of temperatures with elevation]. For completeness the power spectrum of absolute temperature is computed in Appendix \ref{app:avg_temp}. However, subtracting the underlying mean, we decouple these systematic shifts and thereby examine only the structure of dynamical temperature fluctuations caused by kinematics of the atmosphere. Thus, we follow the convention of utilizing anomalies, which has been shown to more accurately describe climate variability over larger areas \cite{temperature_data}. 

Importantly, the coverage of the grid is a dynamic quantity ranging from 57 \% in 1850 to 99 \% coverage in 2015. Limited coverage can induce a bias in the estimated spherical harmonical decomposition \cite{Ahlers2016,Wieczorek2018}. However, the decomposition in this analysis is robust within this range in coverage, as the overall trends remain unchanged when limiting the analysis to the coverage used in 1850 (see Appendix \ref{app:robustness_check}). Additionally, the significant increase of active recording stations could bias any temporal trend of large modes as more stations may resolves features on finer scales. From 1850 to 1960 the number of active land-temperature stations and sea-temperature observations increased 60-fold and 20-fold respectively (see Appendix \ref{app:num_sta}); Over the same period only a slight evolution of the power-spectrum is observed. In contrast, from 1960 to 2020 the number of active land-temperature stations and sea-temperature observations stayed relatively consistent, with an increase of 15\% and 20\% respectively (see Appendix \ref{app:num_sta}). However, in this period a drastic growth is observed within the power-spectrum. We emphasise, that increasing the number of observations by orders of magnitude still yields consistent power-spectra, so the slightly increased statistics of the last 60 years does explain the recent and drastic evolution of the power-spectrum. For completeness the paper presents the decomposition over all 171 years, but the emphasis remains on the recent evolution from 1960 to 2021, where the coverage remains above 95 \% at all times. Furthermore, this limited period mitigates the potential historical biases in SST from the early 20th century \cite{Chan2019}.

The uniform gridding of the Berkeley Earth Land/Ocean Temperature Record is achieved through an interpolated field with a kriging-based approach [for detail on implementation see \cite{temperature_data}]. Such interpolation will potentially introduce a bias by smoothing structures on the spatial scales of interpolation. However, the coverage grows drastically and in extension interpolation-lengths diminish from 1850-1960. If this bias was significant within this data-set the power of intermediate and high modes should increase from 1850 to 1960 as the suppression of large interpolation-lengths diminishes. Nevertheless, intermediate and large modes [eg. $C_{20}(t)$, $C_{50}(t)$ or $C_{80}(t)$] show no consistent growth throughout this period. From 1960-2020 the sampling over land is dense containing approximately 20.000 active measurement sites with typically around 100 km between stations. A bias from smoothing on scales below 100 km will not significantly affect the planetary or mesoalpha scales analysed below (see Appendix \ref{app:interp}). In contrast, the sampling over sea is relatively sparse with up to 500 km between grid-cells in HadSST; These interpolation-scales could theoretically suppress modes of degree $l>40$. However, note this suppression is not observationally suggested from the power-spectrum itself with modes $l=90$ following the temporal trends and spatial structure of degrees well-outside the potentially suppressed regime (eg. $l=20$). This article provides all spherical harmonics to spatial scales of 200 km, but we caution against any interpretation of structures (especially over sea) below 500 km. 

In addition to the data-resolution differences between land and sea, there exist well established kinematic and dynamical differences between land and sea \cite{land_temp,Kennedy2019}. Therefore, we also present the relative contribution of sea and land temperatures (see Appendix \ref{app:div}). Here we show that land-temperatures (despite the smaller surface area) are the dominant contributor to the shape of the power-spectrum. 

Lastly, it is emphasised that the recorded findings presented remain robust across any coarser grid and any temporal resolution from the scale of decades to days. The monthly time-resolution was chosen due to its global coverage (including both land and ocean temperatures). Furthermore, the presented results are insensitive to the specific choice of analysing The Berkeley Earth Land/Ocean Temperature Record with other coincident catalogues yielding a similar functional form of the power spectrum with a similar evolution across seasons and decades. In Appendix \ref{app:ERA5}, we present a similar power-spectral decomposition of the ERA5 data-set.  


\subsection{Decomposition Accuracy}
The spherical harmonical decomposition is computed with the well-documented and readily accessible pyshtools-software \cite{Wieczorek2018}. With increasingly detailed coverage the estimated Spherical Harmonics decomposition is ensured to converge towards the observed temperature-landscape. However, for any limited spatial resolution the spherical harmonics decomposition remains an approximation of the underlying landscape. In Fig. \ref{fig:l_comb_mol_1} we show an example of a decomposition of temperature anomalies to illustrate the high accuracy of the approach. Utilizing all spherical harmonics up to degree $l=90$, does provide an excellent estimation of the underlying temperature-field. Typical deviations between observations and the spherical harmonic reconstructions (using up to degree $l=90$) are 0.05 K. Furthermore, the observed and reconstructed temperature anomalies are not dominated by systematic trends with altitude or latitude. Such effects are decoupled from the investigated signal because we utilise the temperature anomalies and not the absolute temperature. 

\section{Results}
\label{sec:results}

\subsection{The Lowest Modes}
\label{sec:lowest}

\begin{figure}[ht]
\centering
  \includegraphics[width=\linewidth]{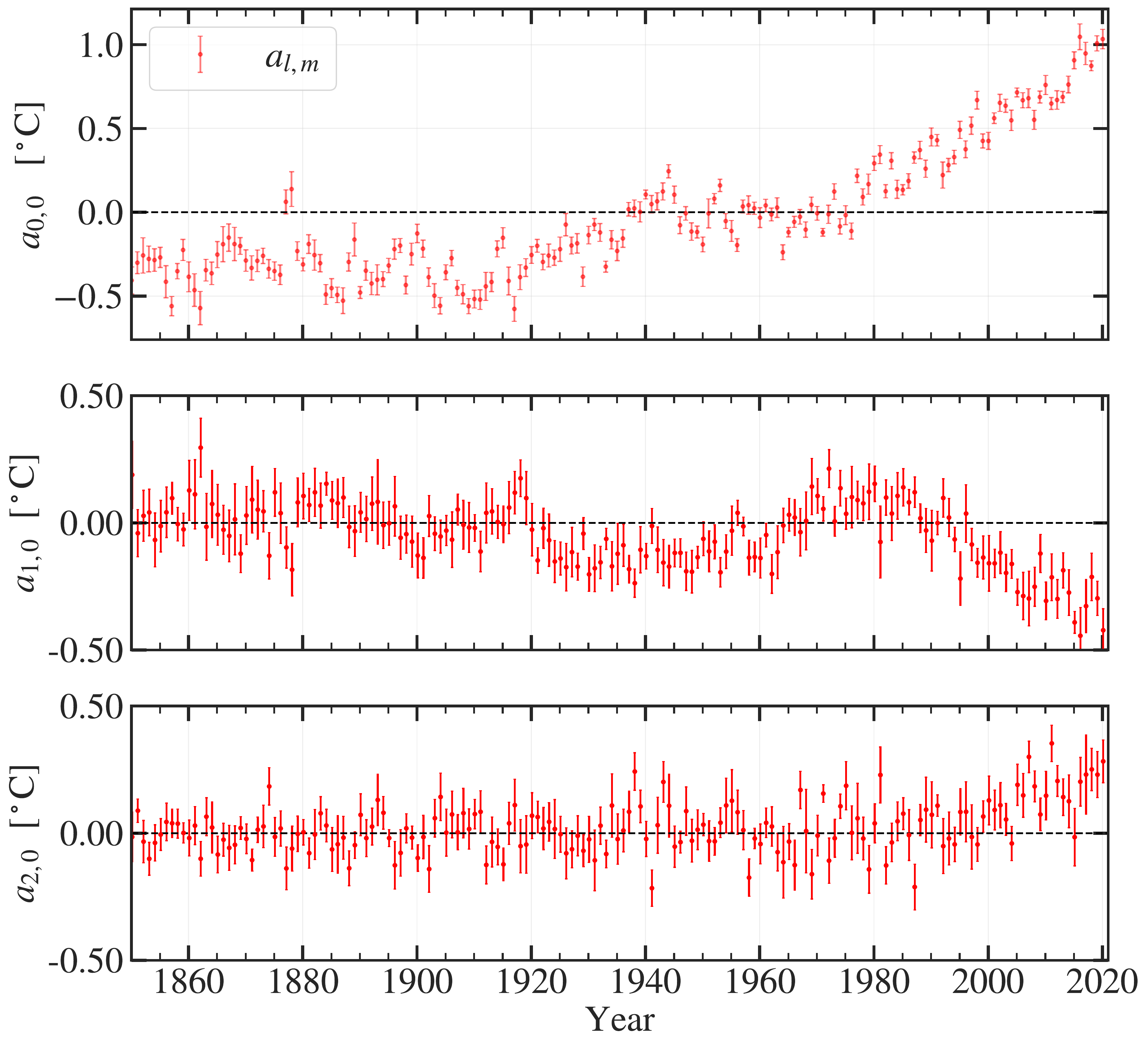} 
  \caption{ $a_{0,0}, a_{1,0}, a_{2,0}$ coefficients of the spherical harmonics derived from global temperature anomalies resolved from Jan 1850 - Jan 2021.} 
\label{fig:temp_low_l}
\end{figure}

With spatially resolved temperatures, we can determine the weights, $a_{l,m}(t)$, of the spherical harmonic functions for the scalar temperature field. To exemplify the information available in the time-series of spectral weights, we show the lowest degree coefficients from 1850 to 2021 in Fig. \ref{fig:temp_low_l}. Each of these time-series contain unique perspectives on the spatial distribution of temperature anomalies, which will now be shortly summarized. \newline

\begin{figure*}
\centering
  \includegraphics[width=\linewidth]{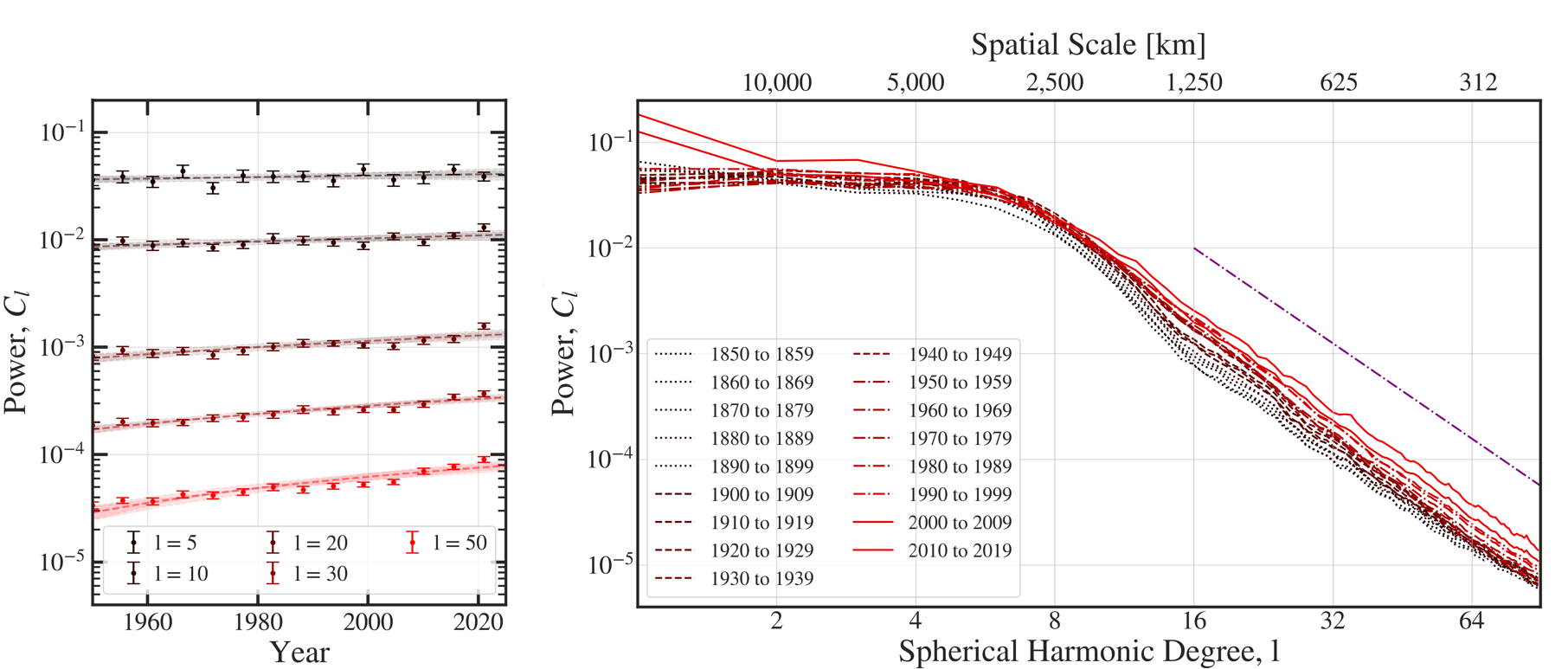}
  \caption{ Right: Power of select modes from 1950-2020 with linear fit overlayed. No systematic temporal evolution of power is seen for the mode l=5, but the power on smaller spatial scales is increasing towards the present. Left: Power spectrum of temperature distribution from 1850s (in black) to 2010s (in red). Dashed-dotted purple line indicates a generic powerlaw with powerlaw index $\alpha=-3$. The power of modes $l \geq 10$ (ie. on spatial scales smaller than 2000 km) is increasing with time especially from 1960 to 2020. The identical power-spectrum of the ERA5-catalogue (which has finer spatial resolution but less temporal range) is shown in Appendix \ref{app:ERA5}.}
\label{fig:power_temp}
\end{figure*}

\textbf{Global Temperature $(a_{0,0})$}: First, the lowest degree (i.e. the spherically symmetric mode) increases towards the present. This mode's evolution demonstrates the increase of the global average temperature \cite{temperature_data}. \newline


\textbf{The North-South Inequality $(a_{1,0})$}: The $Y_1^0(\theta, \varphi)$ spherical harmonic describes variations between northern and southern regions of the sphere. Here $a_{1,0}<0$ implies that the increase in temperature is larger for the Northern Hemisphere than for the Southern Hemisphere. Therefore, the evolution of $a_{1,0}(t)$ reaffirms the well-known interhemispheric temperature asymmetry \cite{Friedman2013}. In the new millennium the Northern Hemisphere has experienced a majority of the temperature increase with the annual interhemispheric temperature asymmetry being approximately 0.4 \degree C in 2010 \cite{temperature_data}. \newline



 

\textbf{ Increased Heating of Polar Regions $(a_{2,0})$}: The $Y_2^0(\theta, \varphi)$ spherical harmonic describes variations between polar and equatorial regions of the sphere. Thus, the growth of $a_{2,0}$ in recent decades validates the well-known polar amplification \cite{Budyko1969,Goosse2018}. Note, the north-south symmetry of $Y_2^0(\theta, \varphi)$ means that $a_{2,0}$ does not constrain the asymmetry of polar amplification \cite{Salzmann2017}.

\subsection{The Power Spectrum}
\label{sec:power}

While the lowest degrees of spherical harmonics remain easily interpretable, the entire power spectrum is required for a comprehensive understanding of the spatial structure. In Fig. \ref{fig:power_temp} the power-spectrum is shown for each decade from 1850 to 2020. 


The overall shape of the power-spectrum is unchanged across decades; the power is constant on the largest of scales, but at a turnover length-scale transitions to a powerlaw-like decline. These two different regimes are dominated by different processes.

First, the observed power is approximately flat for $2\leq l \leq 7$ corresponding to a spatial wavelength from $10.000$ to $3.000$ km. These largest scales contain dynamics on the size of global Rossby waves, large scale tropical waves and the Madden-Julian Oscillation \cite{Zhang2013}. Thus, white noise is observed for temperature anomalies on these planetary scales. 

Second, after the turnover higher degree spherical harmonics become progressively less important following a powerlaw like decline, where the power $C_l \propto l^{\alpha}$. For $l>10$ the power of each harmonic $Y_l^m$ decreases with a powerlaw index which is consistent with $\alpha = -3$ for all parts of the time-series. The powerlaw is observed from spatial scales of 2000 km to the spatial resolution of the data-set (ie. order 50 km - see Fig. \ref{fig:ERA5}). 

Notable, several atmospheric fields including humidity, temperature, zonal-, meridional- and vertical winds at varying altitudes of the ECMWF interim analysis have empirically shown a similar scale-free spectra with varying powerlaw slopes \cite{Lovejoy2011}. However, the specific slope $\alpha=-3$ is remarkable in accordance with the analytical prediction of turbulent flows constrained to 2 dimensions. Under such conditions, the expectation is that the energy spectrum scales as $E(k) \propto k^{-3}$, where the wave-number, $k$, is the 2D-spatial analog of the angular scales of $l$ \cite{Kraichnan1967}. Note, this characteristic scaling of $k=-3$ has been observed previously for potential temperature near the tropopause across planetary and synoptic scales \cite{Nastrom1985}. 

An important test of the classical models of 2D/3D turbulence is to examine the spectral energy transfer. For 2D the energy cascades from small to larger scales, while enstrophy is characterised by a downscale transfer. Previous empirical determinations have yielded little conclusive constraints, although 3rd order velocity structure corroborate the downscale nature of atmospheric cascades \cite{Lindborg1999,Cho2001}. Another analytical perspective on the cause of powerlaw-tails in power spectra is discussed in Ahlers2014 \cite{Ahlers2014}. Here Boltzmann's equation is used to derive a coupled set of differential equations relating the power of each mode to the power of neighbour modes, assuming a simple non-linear coupling associated to turbulence. These coupled equations are solved assuming hierarchical generation, where large spatial fluctuations generate fluctuations on smaller scales. This formalism, suggests that a downward cascade of fluctuations on a sphere would produce a power-spectrum of the form $C_l \propto l^{-3}$. 

Regardless of the causal direction of the cascade a generic prediction of cascades is that an increasing amplitude on large scale variations must have a corresponding increase on all smaller spatial scales. Given the observed growth in the power of the lowest modes (eg. $a_{1,0}, a_{2,0}$) we must therefore expect an increasing power across the spectrum with time. Furthermore, in \S \ \ref{sec:Seasonal}, \S \ \ref{sec:anisotropy} and in Appendix \ref{app:cas} we present evidence of a downward cascade with large scale features being driven externally (such as by differential solar heating and variations over latitudes) which turbulence feeds into a range of fluctuations on all smaller spatial scales. Importantly, this downward cascade predicts that the increasing amplitude of fluctuations on large spatial scales due to climate change, also feeds more volatility for weather on more local scales.





\begin{figure}
    \centering
    \includegraphics[width=\linewidth]{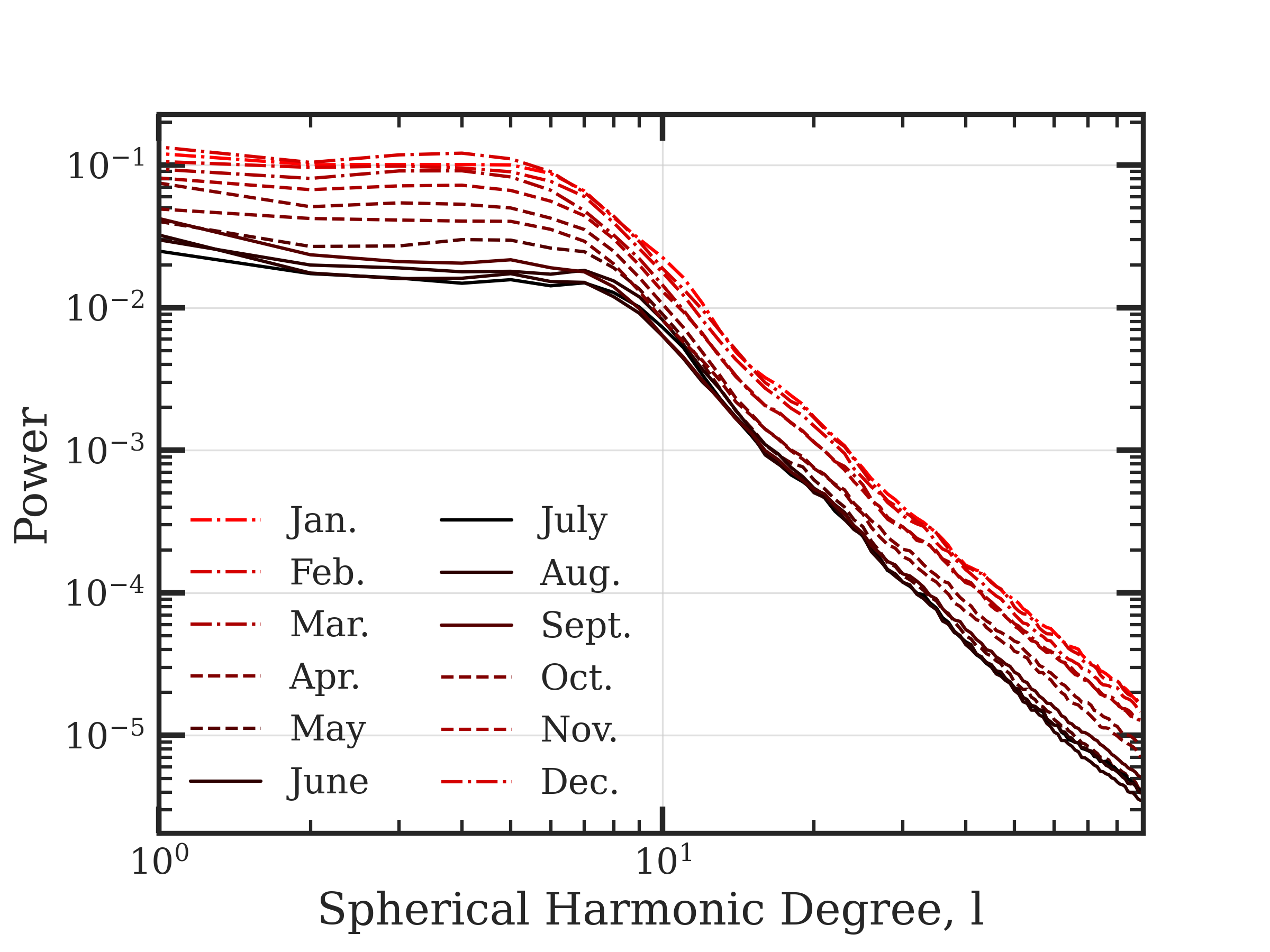}
    \caption{Power spectrum marginalized over month on northern hemisphere. The amplitude of temperature fluctuations on all spatial scales are larger in winter than in summer. Naturally, for the southern hemisphere the order is inverted with most power in June and July. }
    \label{fig:season}
\end{figure}

\subsection{Yearly Evolution of The Power Spectrum }
\label{sec:time_evol}

From 1850 to 1960 the power spectrum shows variability but no systematic coherent growth, but a few noteworthy features emerge after 1960 (see Fig. \ref{fig:power_temp}). First, the power of the lowest degrees is increasing towards the present as discussed in \S \ \ref{sec:lowest}. Second, the flat part of the power-spectrum (degrees $2 \leq l \leq 8$) remains unchanged up to statistical fluctuations. Third, the turnover length-scale increases, ie. the powerlaw cascade begins at lower modes (for elaboration see appendix Fig. \ref{fig:Turnover}). An analogous shift of the energy spectrum to larger spatial scales is often correlated with increased eddy activity and weather volatility \cite{Straus1999}. Lastly, the power of higher degrees ($10 \leq l$) are increasing on all scales from 2000 km to the resolution of the data-set (ie. order 50 km for ERA5) from 1960 to the present. Local temperature fluctuations are growing in magnitude and the growth is accelerating over the past century. 


\subsection{Seasonal Evolution of The Power Spectrum}
\label{sec:Seasonal}
It is observationally well-established that the kinetic energy in the atmosphere is larger in winter than in summer \cite{Straus1999}. In summer the meridional temperature gradient is both weaker and shifted poleward, resulting in a lower eddy activity and a shift to smaller spatial scales. These seasonal features are also seen in the power-spectrum of temperature fluctuations (see Fig. \ref{fig:season}). Summer-months shows decreased fluctuations on all scales (ie. smaller eddy activity), while the transition to the powerlaw decline is shifted to higher wave-numbers (for further elaboration see Appendix \ref{app:turnover}). Importantly, this indicates that variations on small spatial scales follow the growth and decay of fluctuations on the largest scale (ie. that we are observing a \textit{downward} cascade). 

\begin{figure*}
    \centering
    \includegraphics[width=0.95\linewidth]{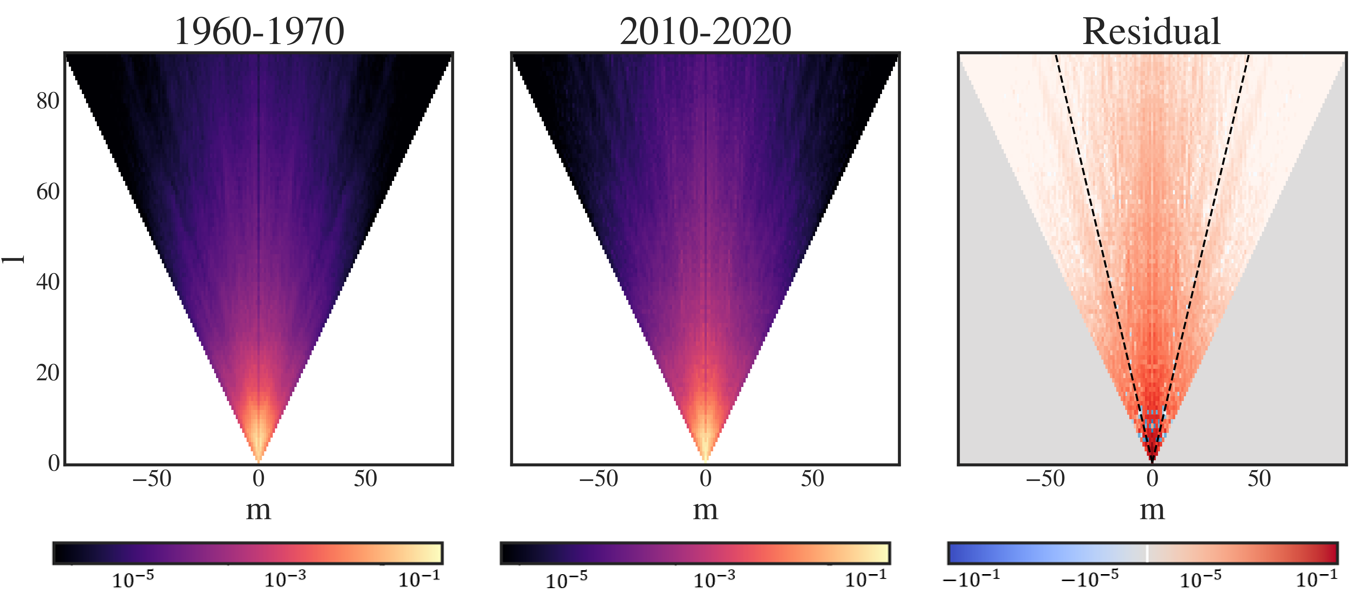}
    \caption{Color indicates power of the $Y_l^m$'th spherical harmonic for the period 1960-1970 [left] and 2010-2020 [middle]. For both periods the power is calculated each month and then averaged over a decade. Right shows residual, suggesting that the increasing power of higher order modes is not symmetric across $m$, but dominated by the increasing power of modes from north-to-south [ie. $m = - l/2$ to $m = l/2$, indicated with dashed lines].}
    \label{fig:power_m}
\end{figure*}

Beyond the yearly fluctuation no other periodicity is detected across the power-spectrum. Therefore, the power of modes on smaller spatial scales only oscillate seasonally (combined with the previously mentioned growth over decades). While the mean global temperature is a complex convolution of anthropogenic climate change and non-anthropogenic oscillations \cite{Wang2019}, the fluctuations from the mean (described by smaller modes) suggests a singular temporal trend. Thus, the power-spectrum yields a diagnostic with a simple and predictive evolution across time.    

\subsection{The Horizontal Anisotropy of Climate Change}
\label{sec:anisotropy}
The expectation from atmospheric reanalysis fields is that horizontal anisotropies are common, with fluxes elongated more zonally than meriodionally \cite{Lovejoy2011}. The easily interpretable decomposition of spherical harmonics does provide a straightforward framework for examining horizontal anisotropies, but by summing over $m$ we have yet to discuss the spatial orientation of these fluctuations. In spherical harmonics, spherical symmetry implies every $m$ should contribute equally to the power of the $l$'th mode. This is elaborated through simulated temperature fields in Appendix \ref{app:sim_ani}. Here it is shown that when generating temperature fluctuations with isotropic variances along latitudes and longitudes the power is uniform across $m$. However, if the length-scale of temperature fluctuations is smaller along longitudes than latitudes the modes from $m \approx -\frac{l}{2}$ to $m \approx \frac{l}{2}$ are shown to be the dominant contribution to the power. 

Indeed, anisotropy is observed throughout the timeseries as seen in Fig. \ref{fig:power_m} [left and middle panel]. Evidently, the climate generically has a larger magnitude of temperature fluctuations meridionally than zonally, as expected for a rotating sphere \cite{Lovejoy2011}. 
However, in contrast to previous analyses the detailed timeseries of global temperatures allows computing the change in power across decades [right panel]. We see that on all spatial scales resolvable it is mainly fluctuations oriented north-to-south which are increasing in amplitude. This coherence is a prediction of downward cascades, where the observed growing fluctuation on large scales (eg. $a_{1,0}, a_{2,0}$, etc.) feed increasing amplitudes on all subsequent scales. Because the large scale fluctuations are oriented north-to-south the cascade displays a similar anisotropy. In contrast, an inverse cascade should not display any horizontal anisotropy on small scales, as there are no local differences between latitude and longitude. Climate change therefore results in rapid weather transitions which are growing in frequency and magnitude, but crucially the increase is primarily driven by increased volatility along longitudes. 


\section{Conclusion}
Spherical harmonics provide a succinct description of the Earth's climate. Within this paper we prove that the lowest 90 modes can reconstruct global temperature anomalies with a typically accuracy of 0.05 K. This highly accurate decomposition is the result of a simple combination of oscillations on different spatial scales, which is succinctly summarized by the shape of the power-spectrum. 
The largest power is coming from modes $l \leq 8$, with progressively smaller scales contributing less and less. Importantly, the decline in power on spatial scales from $2.000$ to $50$ km follows a powerlaw with exponent $\alpha=-3$, which is characteristic of 2 dimensional turbulence and other theoretical models of cascading fluctuations. Furthermore, mapping the seasonal fluctuations and anisotropies between longitudes and latitudes we show that large scale temperature variations cascades into fluctuations on all smaller spatial scales. 

As previously observed large scale fluctuations (which are oriented from north-to-south) are growing in magnitude. Therefore, the observed cascade dictates that on any scale the amplitude of fluctuations along longitudes has to increase. From the planetary scale to that of weather fronts, temperature oscillations are becoming more volatile due to climate change. Ultimately, the methodology provides an illuminating perspective on our evolving climate by bridging the gap between known large scale structures of the climate and small scale volatility.

Furthermore, the methodology introduced within the paper can be extended with significant theoretical and observational benefits. 
First, larger spatial resolution will allow improvements on the accuracy of reconstructions by decomposing the structure at even smaller spatial scales. The power structure of higher modes could reveal the extent and the small-scale limit of the observed cascade; a more detailed grid could even examine the regime of 3-dimensional turbulence. Second, larger temporal resolution allows a closer examination of the flow of fluctuations, i.e. how the spatial scales are causally connected. Thereby, probing the underlying physics connecting the cascade (discussed in Appendix \ref{app:cas}). Lastly, we emphasise that the observed power-spectrum is remarkable simple. Three parameters, a normalisation, a turnover-length and a powerlaw-slope, characterize the entire power spectrum of temperature anomalies for over 170 years of observations from the planetary scale through the mesoalpha. This remarkable universality should be examined, reproduced and predicted by future climate models in order to further constrain the spatial scales of climate change. \newline



\section*{Acknowledgements}

The author would like to thank Mogens Høgh Jensen, Peter Ditlevsen, Markus Ahlers and Rune Thinggard Hansen for useful discussions on drafts and insightful feedback. Furthermore, Kevin Kumar, Gergely Friss, Gustav Ernst Madsen, Rasmus Damgaard Nielsen and Jason Koskinen all deserve praise for kind encouragement with the conceptual formulation. Lastly, Julie Kiel Holm provided the audacity required for publication. The Cosmic Dawn Center (DAWN) is funded by the Danish National Research Foundation under grant No. 140. \newline   

\section*{Data Availability Statement}
The data was collected from The Berkeley Earth Land/Ocean Temperature Record, which can be accessed at \url{http://berkeleyearth.org/data/}. 

\bibliography{sample}



\setcounter{equation}{0}
\setcounter{figure}{0}
\renewcommand{\thefigure}{A\arabic{figure}}

\begin{figure*}
    \centering
    \includegraphics[width=\linewidth]{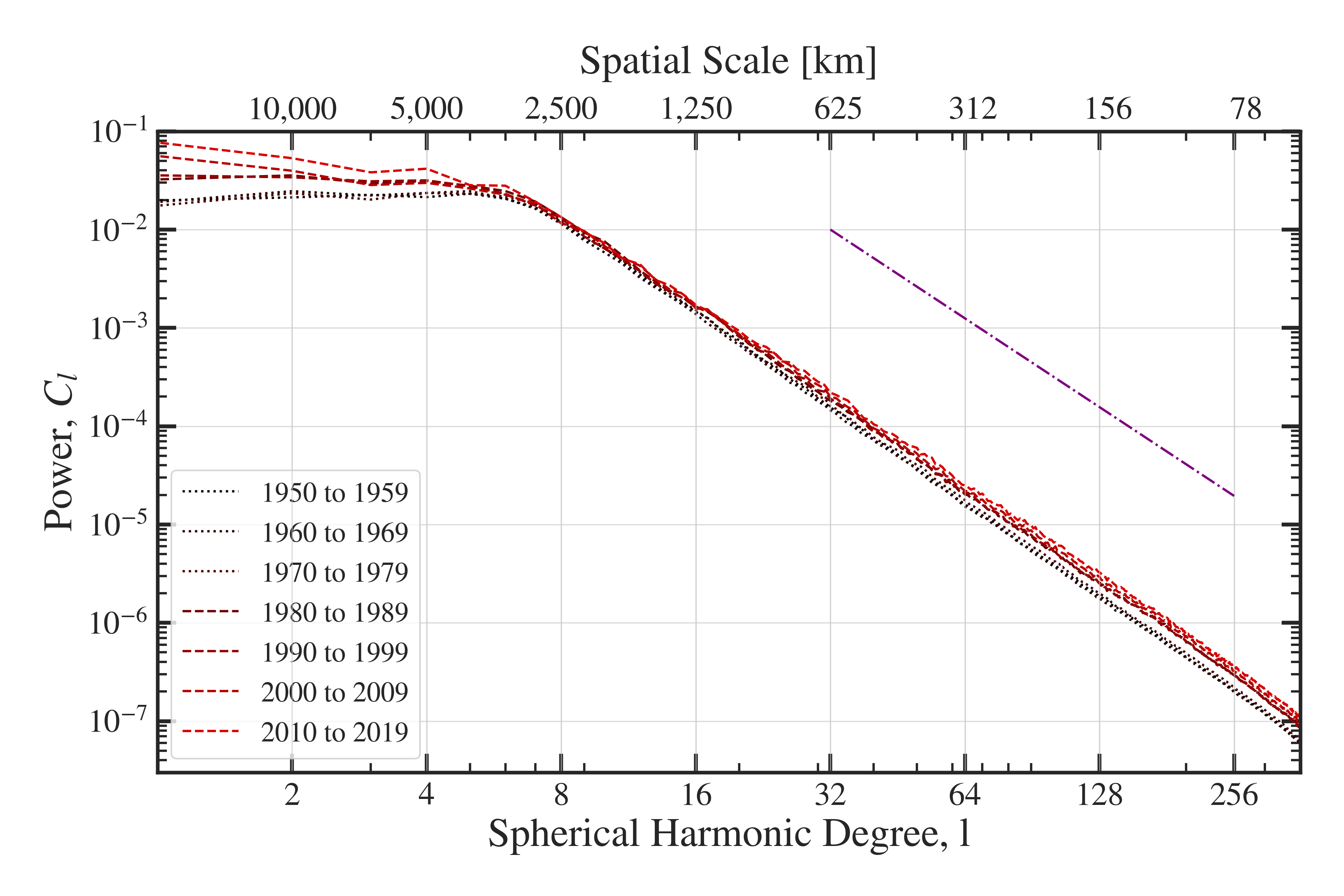}
    \caption{Left: Power spectrum of temperature distribution from 1950s (in black) to 2010s (in red) from ERA5 catalogue. Dashed-dotted purple line indicates a generic powerlaw with powerlaw index $\alpha=-3$. The power of modes $l \geq 10$ (ie. on spatial scales smaller than 2000-50 km) is increasing with time.} 
    \label{fig:ERA5}
\end{figure*}

\newpage
\appendix
\section*{Appendix}
\renewcommand{\thesubsection}{\Alph{subsection}}

\subsection{Power Spectral Analysis of ERA5 Dataset}\label{app:ERA5}
While the above analysis centers on The Berkeley Earth Land/Ocean Temperature Record - other coincident catalogues tell a similar story. In Fig. \ref{fig:ERA5}, the power spectrum for each decade from temperature anomalies in the ERA5 reanalysis (from 1979 to present) and the preliminary ERA5 reanalysis (from 1950 to 1978) is illustrated. The anomalies are defined relative to the locally inferred average temperature for that month over the period Jan 1951-Dec 1980 (ie. similar to the Berkeley Earth Land/Ocean Temperature Record). Crucially, this shows an identical functional form of the power spectrum with a similar normalization, turnover length-scale and powerlaw-decline (see Fig. \ref{fig:power_temp}). Additionally, the evolution with time displays a coherent growth of power for modes within and above the turbulent cascade. 

Note, the analysis of the 30km spatial resolution of ERA5 allows an examination of small modes ($90<l<360$). The previous observed powerlaw-regime with $\alpha=-3$ is observed within the entire range from 2000 to 50 km. The scale-height of the atmosphere (order 10 km) still remains significantly smaller than the horizontal resolution, so at first glance one might believe we are still well above the scales of 3-dimensional turbulence, but several previous analysis (especially of wind spectra within the Global Assimilation and Prognosis System) indicate $\alpha=-5/3$ regime up to several hundreds of kilometers \cite{Nastrom1985,Lilly1989}. Thus, the singular powerlaw extending over several orders of magnitude for temperature anomalies on the earths surface is seemingly in contrast with some atmospheric fields at higher altitudes. Nevertheless, we emphasise that ERA5 reanalysis is not an empirical field - it is obtained through variational data assimilation techniques, which require smoothing over small spatial scales. Thus, the spectral power on the smallest scales remains ill-constrained is subject to the potential systematics effects of smoothing (see Appendix \ref{app:interp}).  


\subsection{Power Spectrum of Average Temperature}\label{app:avg_temp}
\begin{figure}[h!]
    \centering
    \includegraphics[width=\linewidth]{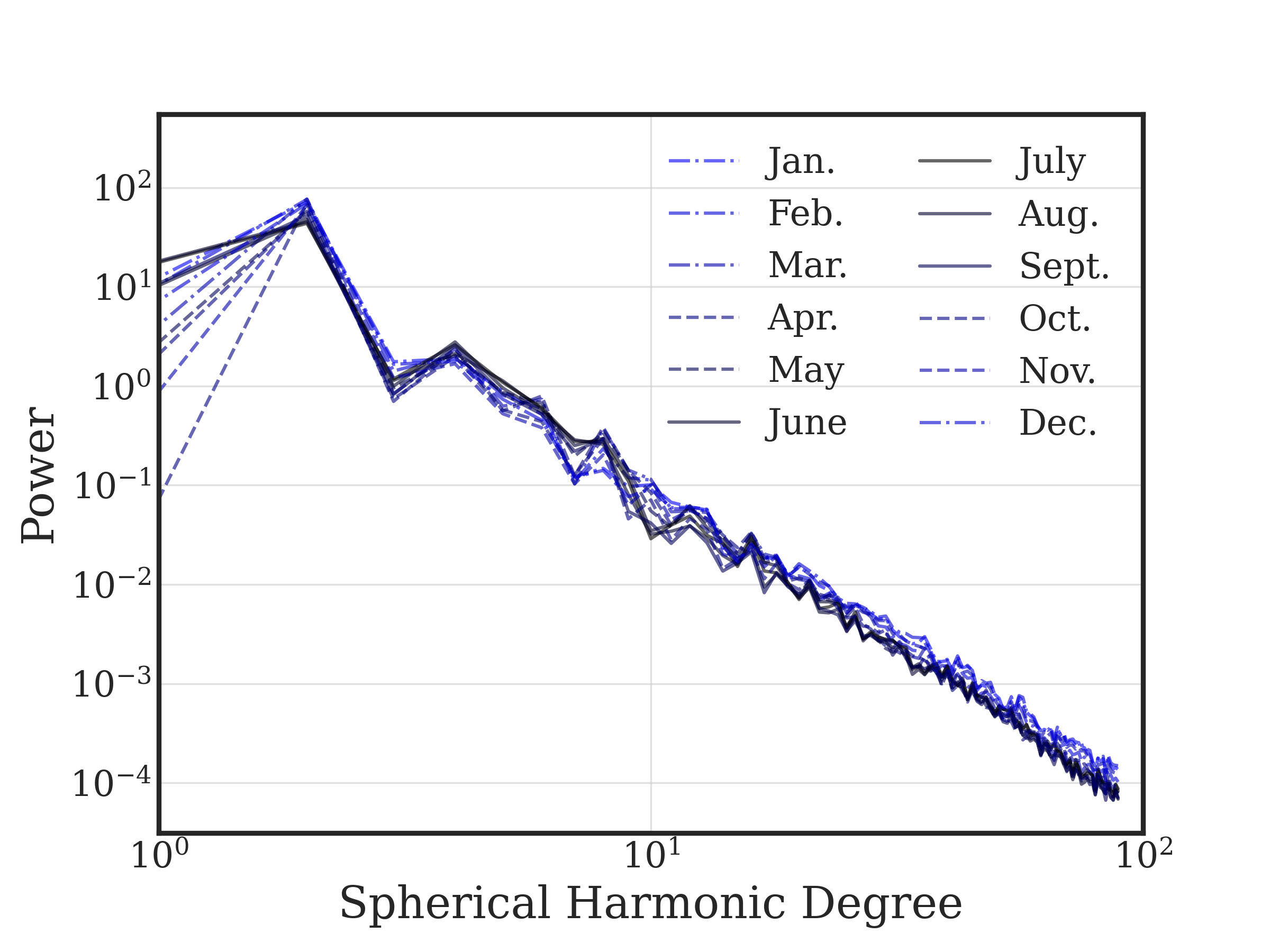}
    \caption{Power spectrum of average temperature for each month from Jan. 1951 to Dec. 1980. The power for all modes is larger than for temperature anomalies. }
    \label{fig:abs}
\end{figure}

Temperature anomalies are reported relative to the baseline average temperature for each month from Jan. 1951 to Dec. 1980. The power spectrum for this baseline temperature is seen in Fig. \ref{fig:abs}. Firstly, the power of all modes is larger than observed for temperature anomalies; this is expected as absolute temperature varies far more drastically - ranging from $-70 \degree$C to $40 \degree$C with elevation, climate and latitude. For instance the systematic variation with elevation will increase the amplitude of temperature fluctuations on spatial scales ranging from continents to mountains. Secondly, the shape of the power-spectrum is different than temperature anomalies, but still consistent with a powerlaw for large degrees. The powerlaw  slope $\alpha$ is for all months within the range $-3.1$ to $-3.4$, which remains significantly steeper than for temperature anomalies. The cause of the increased steepness remains (as of yet) unresolved. Thirdly, the spread from the powerlaw has increased with large systematic bias observed across months for specific degrees (ie. $C_3$ is remarkably small). Such rugged power-spectra are commonly observed when examining altitude \cite{Wieczorek2005}, which illustrates the apparent systematic bias in decomposing absolute temperature at the surface. 

Spherical harmonics have not previously been used as a standard descriptor of climate change, because structures like elevation, latitude and differences between land and ocean dominate the variation of absolute temperature. However, by limiting the analysis to temperature anomalies we focus only on the underlying fluctuations due to atmospheric processes. This effectively decouples the effect of climate change from the larger irrelevant structures. 

\subsection{Power Spectrum Characteristics are robust to limiting Coverage}\label{app:robustness_check}
\begin{figure}[h!]
    \centering
    \includegraphics[width=\linewidth]{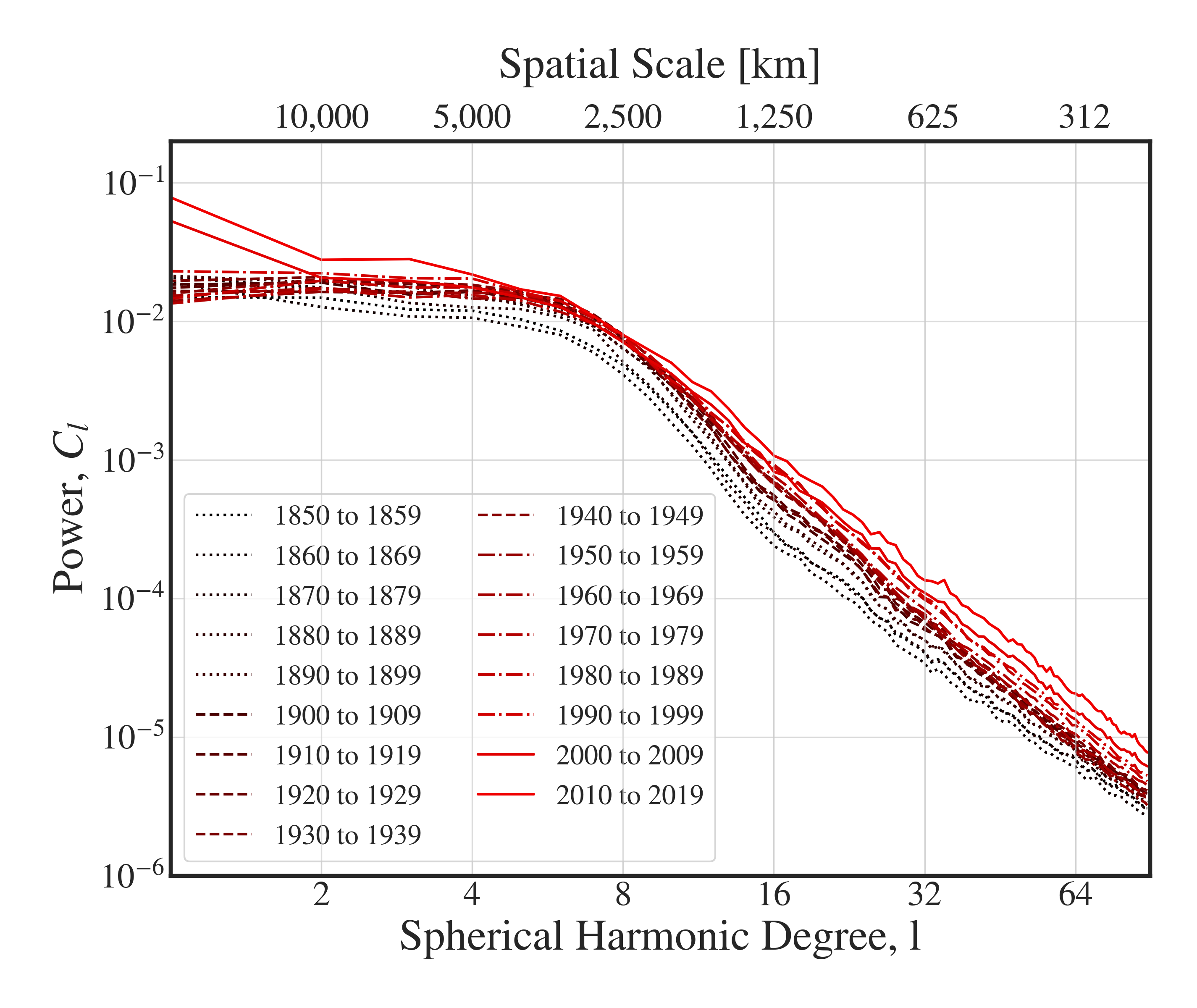}
    \caption{Power spectrum of temperature distribution from 1850s (in black) to 2010s (in red) in log-log plot limiting coverage to that of 1850. The power for modes $l \geq 10$ is increasing with time especially from 1960 to 2020. For comparison see the full coverage in Fig. \ref{fig:power_temp}. \label{fig:coverage}}
\end{figure}

While the specific numerical value of the power for any mode is dependent on the coverage, the characteristic features of the power-spectrum (discussed in \S \ \ref{sec:power} and \S \ \ref{sec:time_evol}) are unchanged when limiting the coverage to that 1850 [See Fig. \ref{fig:coverage}]. The turnover-length remains at $\approx 3000$ km, the powerlaw slope remains consistent with $\alpha=-3$ for the entire time-series and the relative growth of small to large modes is unchanged. 

\subsection{Increasing Resolution}\label{app:num_sta}
\begin{figure}[h!]
    \centering
    \includegraphics[width=\linewidth]{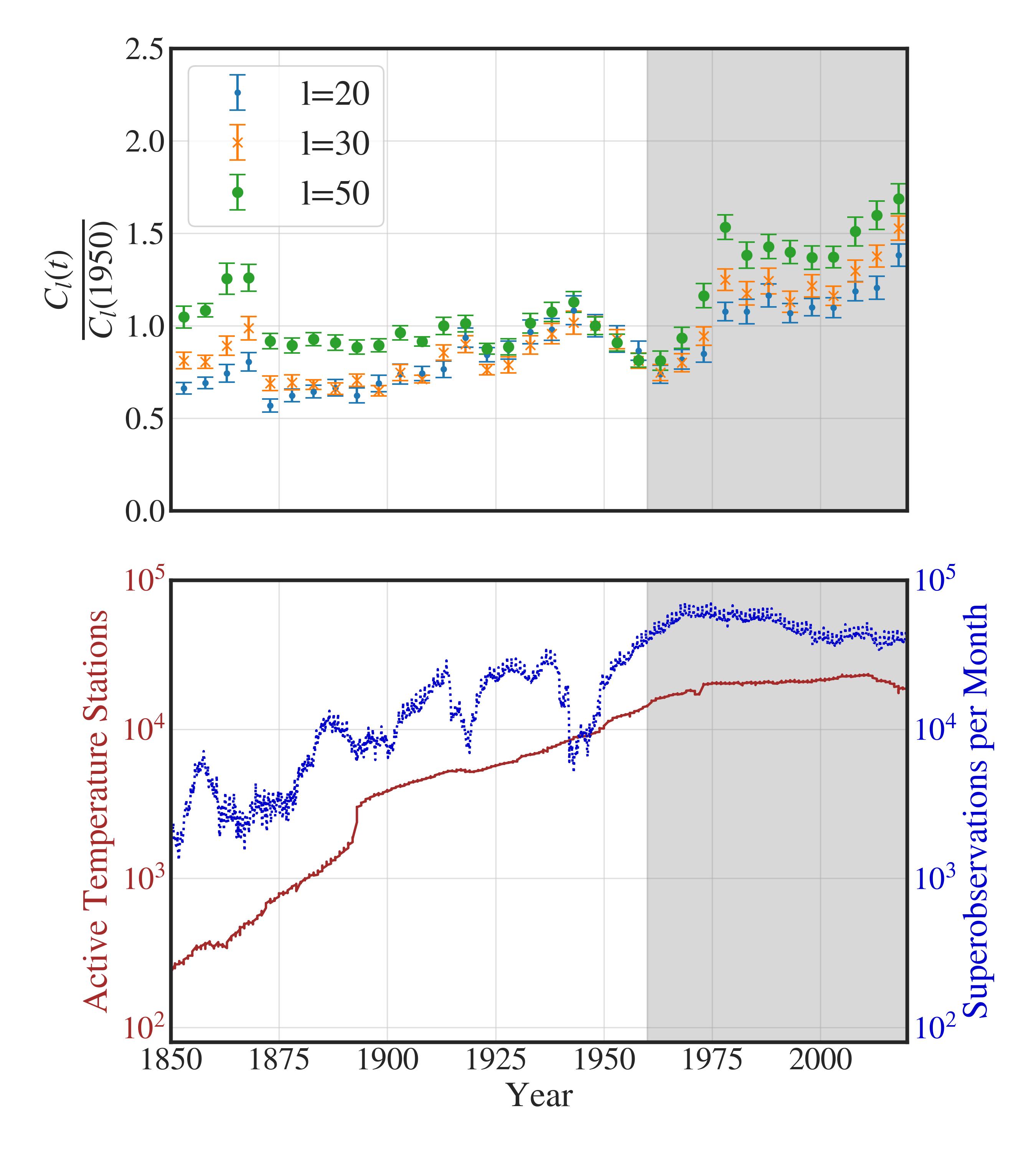}
    \caption{Top: power of degree l relative to power in 1850. Bottom: Number of active temperature stations over land (brown line) and number of "superobservations" of temperatures over sea (blue dots) from 1850-2021. Grey shaded region indicates the period 1960-2021. The significant growth in power-spectrum does not correlate with an increasing number of observing stations.}
    \label{fig:n_obs}
\end{figure}

From 1850-2020 the number of surface temperature measurements over land and sea increased drastically. In Fig. \ref{fig:n_obs} we show the evolution in number of active temperature stations over land (reported in \cite{land_temp}) and number of "superobservations" of temperatures over sea. Super-observations are the Winsorised mean of temperature anomalies used in HadSST \cite{Kennedy2019}.

We emphasise that most of the evolution of the power-spectrum is from 1960-2020, where no significant change in number of either land stations nor super-observations over sea takes place. On the other hand, the methodology yields consistent power-spectra from 1850-1960 even though the number-of-observations increased by orders of magnitude. Thus, increasing the number of measurements will improve the statistical uncertainty on the monthly temperature-estimates, but improving accuracy does not correlate with a bias in the spherical harmonics decomposition. 

\subsection{Interpolation Bias}\label{app:interp}
\begin{figure}[h!]
    \centering
    \includegraphics[width=\linewidth]{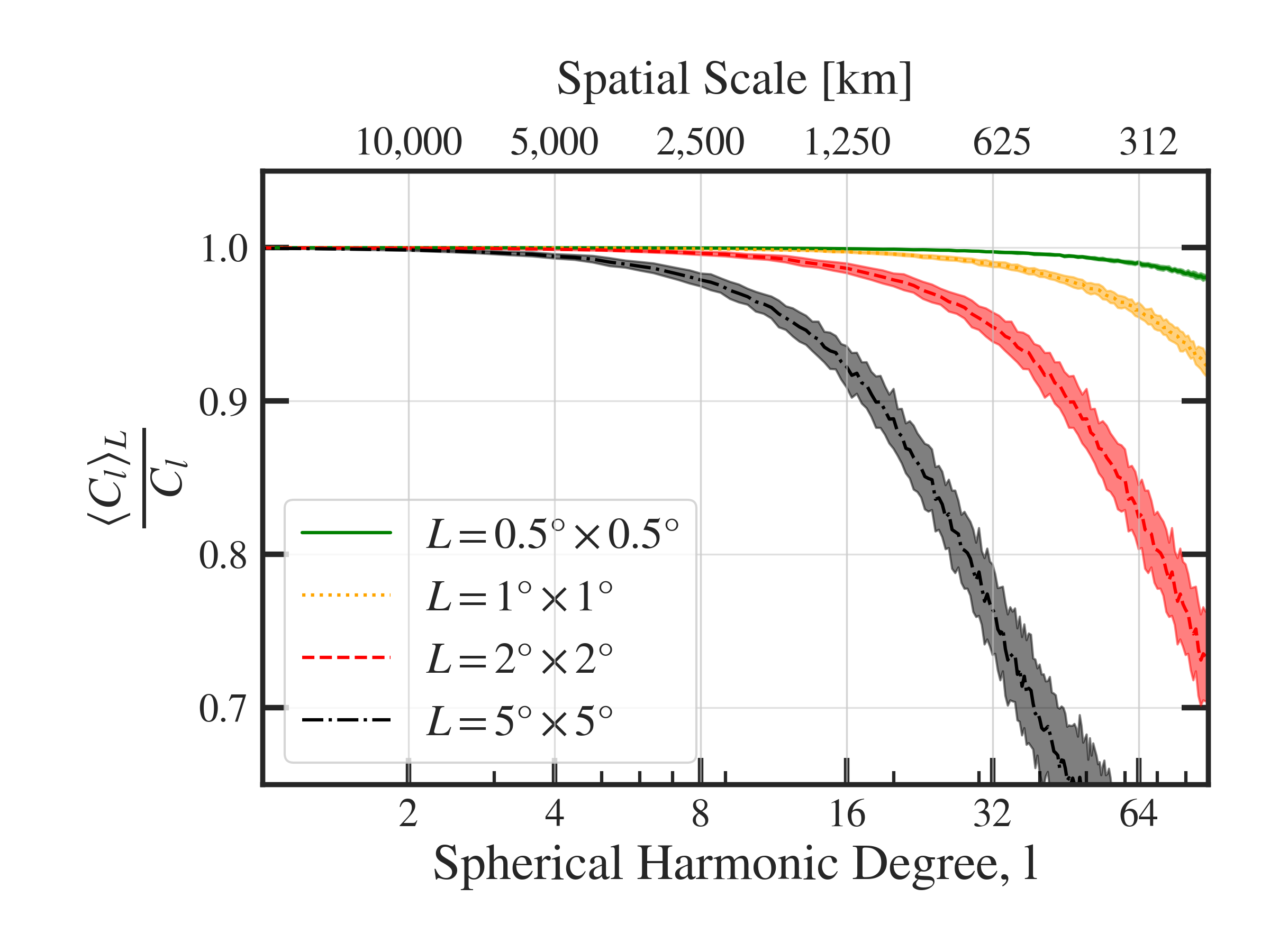}
    \caption{Inferred power for smoothed temperature anomalies (linearly interpolated over 0.5\degree, 1\degree, 2\degree and 5 \degree latitude and longitude) 
    relative to power of unsmoothed temperature anomalies. The smoothed power is underestimated with most suppression at small angular scales. Larger interpolation lengths result in smoothing structure at larger scales.} 
    \label{fig:interp}
\end{figure}

\begin{figure}[h!]
    \centering
    \includegraphics[width=\linewidth]{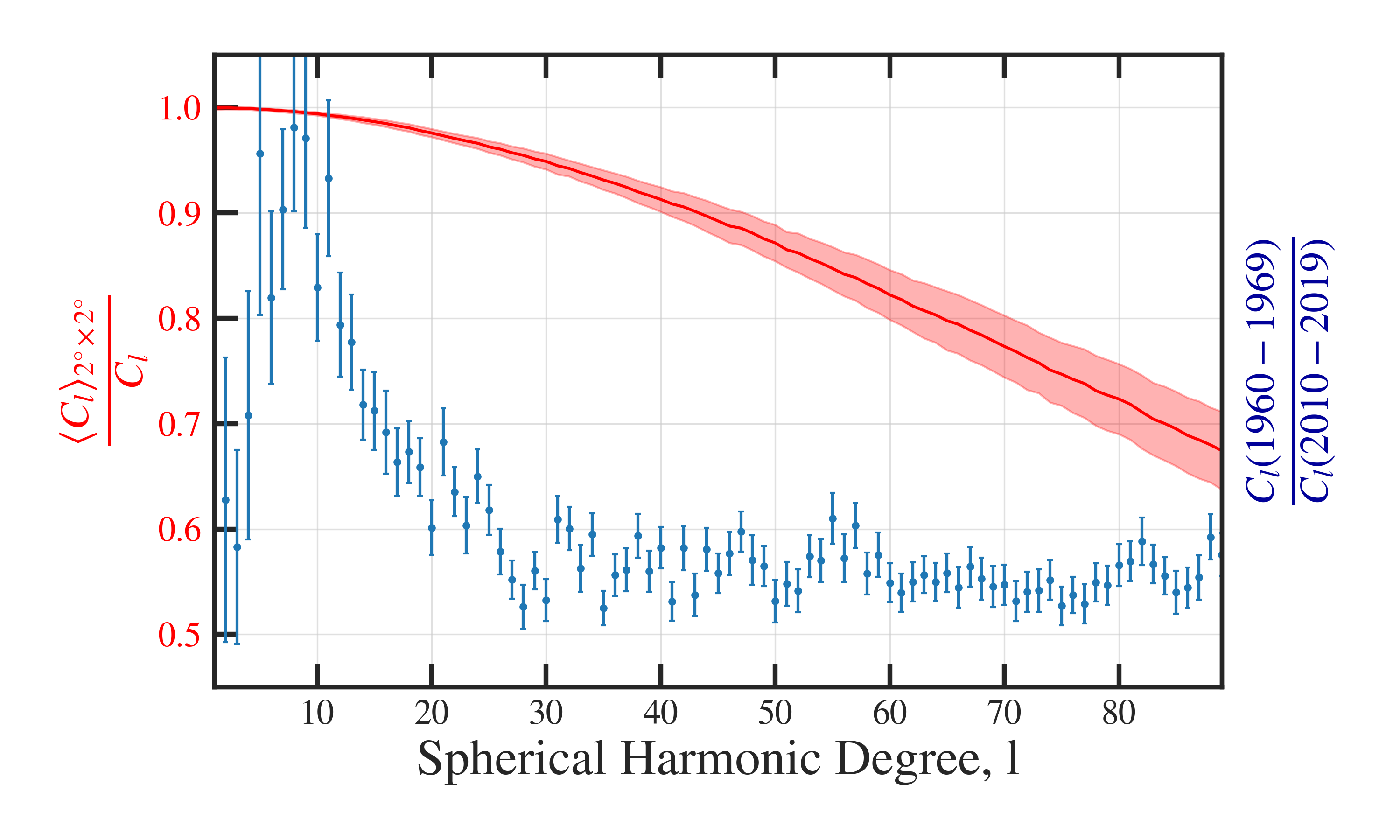}
    \caption{Inferred power for smoothed temperature anomalies (linearly interpolated over $2\degree \times 2 \degree$ lat-lon) relative to power of unsmoothed temperature anomalies (red line). Observed power in 1960 to 1970 relative to power in 2010 to 2020 (blue error-bars). The angular dependence of any interpolation bias contrast the coherent growth of modes $20 \leq l \leq 90$.} 
    \label{fig:interp_1}
\end{figure}

Fig. \ref{fig:interp} shows the underestimation of power when generated temperature anomalies are interpolated over a characteristic length-scale, $L$. As expected larger interpolation length results in a larger bias. For interpolation over $0.5\degree$, $1\degree$ and $2\degree$ latitude and longitude the estimated power remains accurate within respectively 1\%, 3\% and 10\% up to degree $l=50$. Therefore, the typical bias remains significantly smaller, than the factor 2 growth in power of these modes from 1960-2020. Furthermore, the bias has a clear dependence on angular scale. In contrast, from 1960-2020 all power within the cascade grew coherently maintaining the powerlaw-slope, $\alpha=-3$ (see Fig. \ref{fig:interp_1}).

Over land the typical scale of interpolation from 1960-2020 is less than 100 km so any potential bias would remains negligible for modes $l \leq 90$ \cite{land_temp}. Note, the growth observed in the power-spectrum from 1960-2020 is also detected in a when only examines temperature anomalies over land using a multi-tapered power-spectrum.

The HadSST grid-cells are separated by up to 500 km, which is re-interpolated into a $1\degree \times 1\degree$ lat-long grid \cite{temperature_data}. Thus, interpolation lengths are typically of order 250 km (which suggests underestimation of around 20\% for degree $l=50$). Thus, sparse sea coverage may bias the power-spectrum on small scales at all times, but neither the magnitude nor the functional shape of the bias matches the growing power from 1960-2020. 

\subsection{Difference between Land And Sea}\label{app:div}
\begin{figure}[h!]
    \centering
    \includegraphics[width=\linewidth]{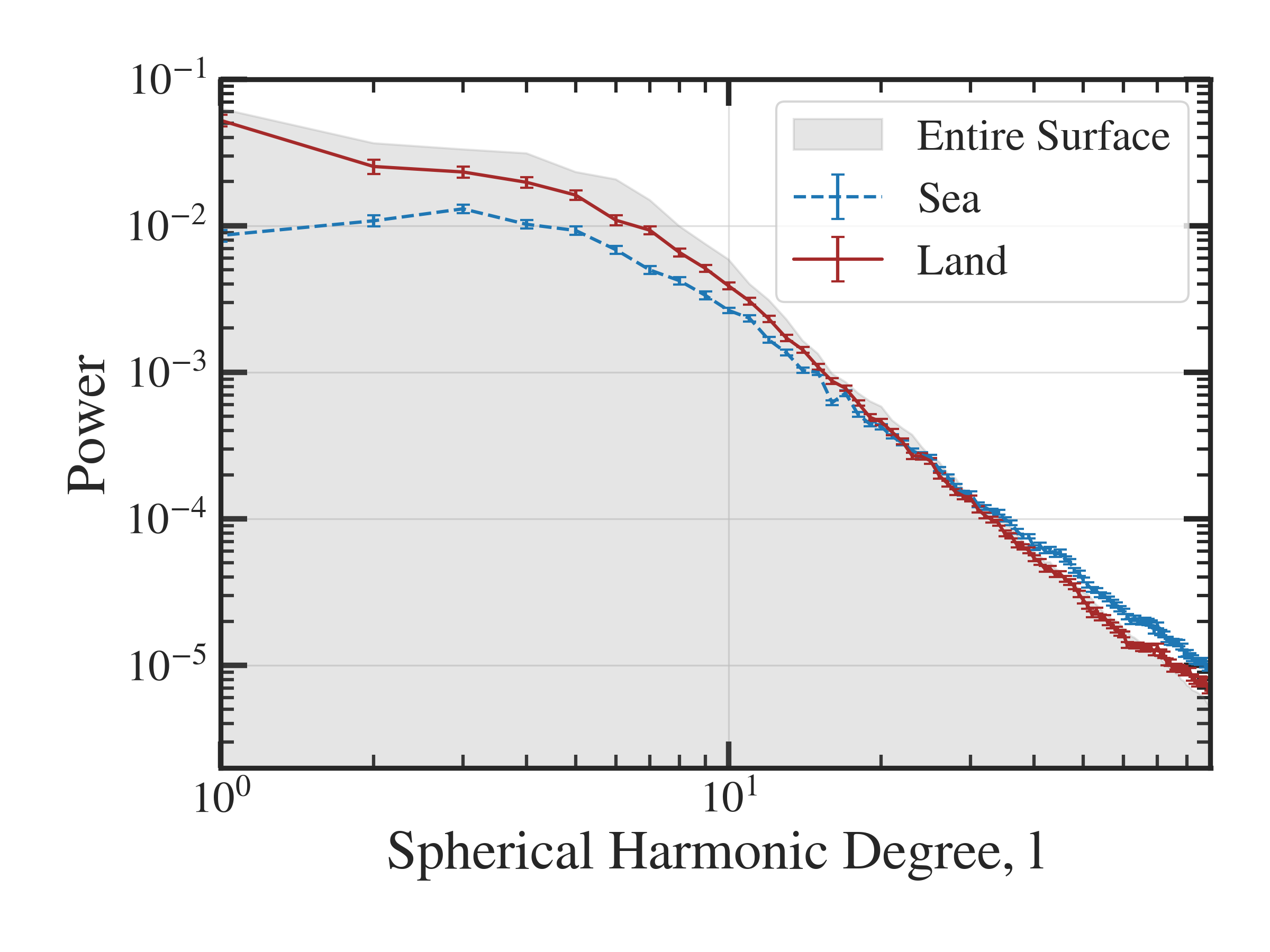}
    \caption{Power spectrum from 2000-2020 when only calculated from Land or Sea temperatures. Importantly, the temperature fluctuations are not consistent with identical within 5$\sigma$. The power spectrum over land is similar to the entire surface (indicated by grey shading) with similar powerlaw slopes, turnover length-scale and overall power on large scales. Sea temperatures display fewer large scale temperature fluctuations, but with a shallower powerlaw decline than land. }
    \label{fig:earth_sea}
\end{figure}

The analysis presented thus far has averaged over the entire globe, thereby neglecting any kinematic or dynamical differences between land and sea. However, if we limit the spatial coverage, we can determine the contribution of different parts of the globe. In general, the procedure of obtaining a multitaper power spectrum estimate is well established. \cite{Wieczorek2005,Wieczorek2018} As seen in Fig. \ref{fig:earth_sea}, the contribution of earth and sea temperatures are not interchangeable with slightly different functional shapes and powerlaw slopes. While the typical temperature fluctuations over land and sea have a similar turnover length, the fluctuation on most length scales have larger amplitudes on land than at sea. For instance the dipole power, $C_1$, is substantially larger over land than sea - as extensively observed the seasonal difference is larger over land. 

We again emphasise that limiting coverage will introduce a bias for degrees close to the spectral bandwidth of the aliased structure \cite{Ahlers2016,Wieczorek2018}. This is a direct consequence of the signal from low degrees leaking into high degrees over the spectral bandwidth of the excluded regions. However, previous studies specifically dividing land and sea have highlighted 1) that any leakage of signal is small between these domains and 2) any bias remains small for large degrees \cite{Wieczorek2018}. Thus, we conclude that the spectral differences of land and sea is not merely caused by aliasing the spatial structure. Furthermore, Fig. \ref{fig:earth_sea} clearly indicates that the land-surface temperature field remains the dominant contribution to the observed power spectrum. The overall power, the turnover length and the powerlaw decline for the entire globe traces the power spectrum from land.  


\subsection{The Flow of Temperature Fluctuations}\label{app:cas}
\begin{figure}[h!]
    \centering
\includegraphics[width=\linewidth]{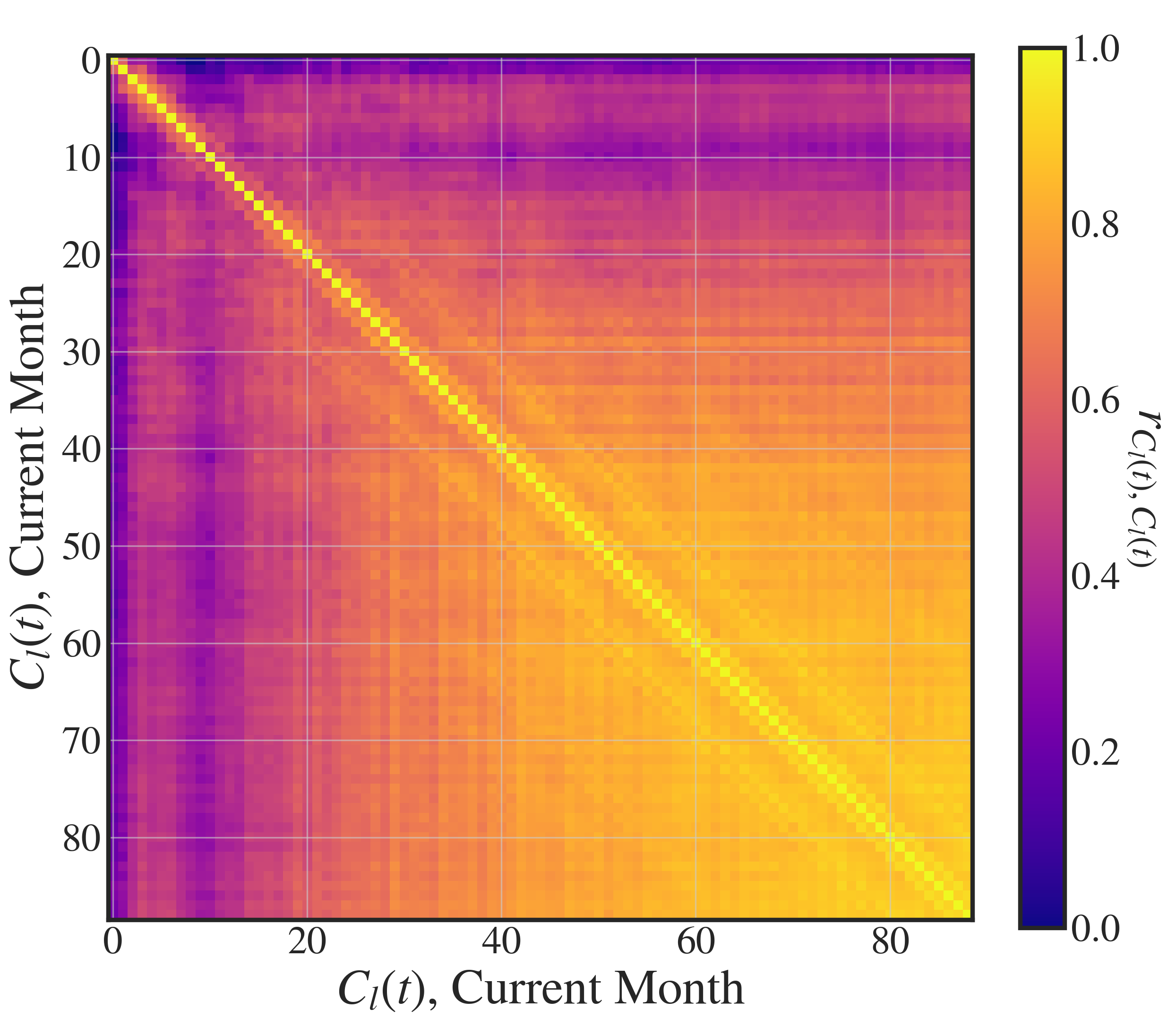}
    \caption{Correlation Coefficient between power of modes. Typically modes of similar spatial scales are more correlated while all modes within the cascade show large correlation.}
    \label{fig:cascade}
\end{figure}

To determine whether the observed cascade is downwards (from large spatial scales to smaller) or inverse (from small to large) we compute the Pearson Correlation Coefficient between the power of each mode; 
\begin{equation}
    r_{C_l(t),C_l(t+\Delta t)} = \frac{Cov(C_l(t),C_l(t+\Delta t))}{\sigma_{C_l(t)} \sigma_{C_l(t+\Delta t)}}
\end{equation}
For $\Delta t = 0$, we can determine the correlation between the power of different modes. In Fig. \ref{fig:cascade} we see that all modes within the cascade ($l \geq 10$) show large internal correlation with $r > 0.9$. The large scale spatial structures are far more uncorrelated with $r<0.6$. 

\begin{figure*}
    \centering
\includegraphics[width=0.95\linewidth]{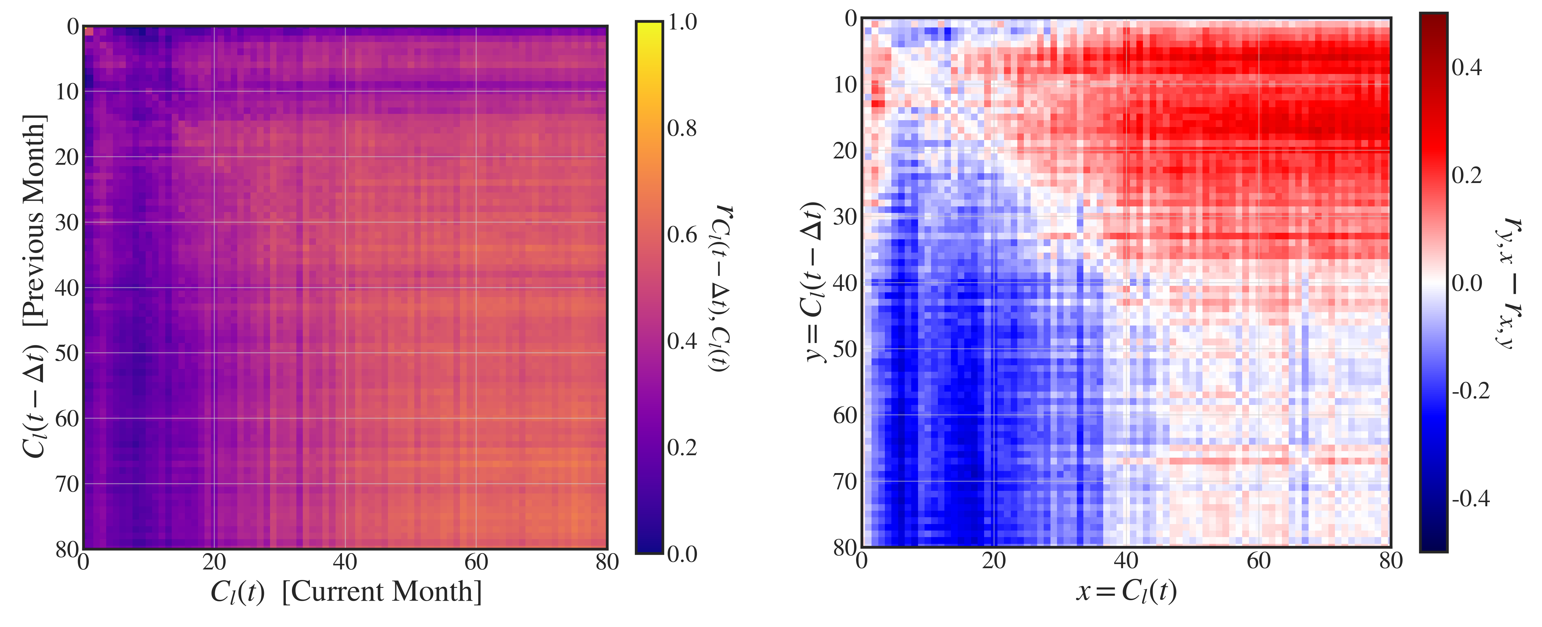}
    \caption{Correlation Coefficient between power of modes in neighbouring months (left) and residual correlation coefficient [ie. difference in correlation between power in prior and subsequent months] (right). We still see that modes of similar spatial scales are more correlated, while all modes within the cascade show large correlation. However, the correlation between powers across subsequent months is not symmetric. There is more correlation between lower modes in previous months with higher modes in subsequent months then between higher modes in previous months with lower modes in ensuing months. }
    \label{fig:cascade2}
\end{figure*}

For $\Delta t = 1$ Month, we can determine the correlation between the power of different modes in neighbouring months, see Fig. \ref{fig:cascade2} (left). Over time correlations in general decay, as the former temperature fluctuations diffuse and dissipate. Nevertheless, the relative distribution of correlations remain unchanged with for instance the highest correlation remaining within the cascade. However, the correlation is no longer symmetrical; for instance $C_{20}(t + \Delta t)$ is more correlated with the power of higher modes in subsequent months, than the correlation between $C_{20}(t)$ and higher modes of prior months. This implies, that increasing power on large spatial scales is often followed by increased power on smaller spatial scales, but the inverse does not hold. Thus, the direction of the cascade of temperature oscillations is downwards. 

In Fig. \ref{fig:cascade2} (right) this contrast is emphasised by taking the difference in correlation between previous and following month. All modes from $10 \leq l \leq 40$ the flow is from large to smaller spatial scales (ie. the correlations from large to small are forward in time). Ultimately, this resolution across time yields observational evidence, that temperature oscillations follow a downward cascade from large to small spatial scales. For modes $l < 8$ the inverse relationship seems to hold, smaller spatial scales precede oscillations on larger scales. This tentatively suggests, that temperature oscillations are generally driven around the characteristic transition within the power-spectrum (ie. $l \approx 8$), which then feeds temperature fluctuations on both smaller and larger spatial scales. 

\begin{figure*}
    \centering
    \includegraphics[width=0.95\linewidth]{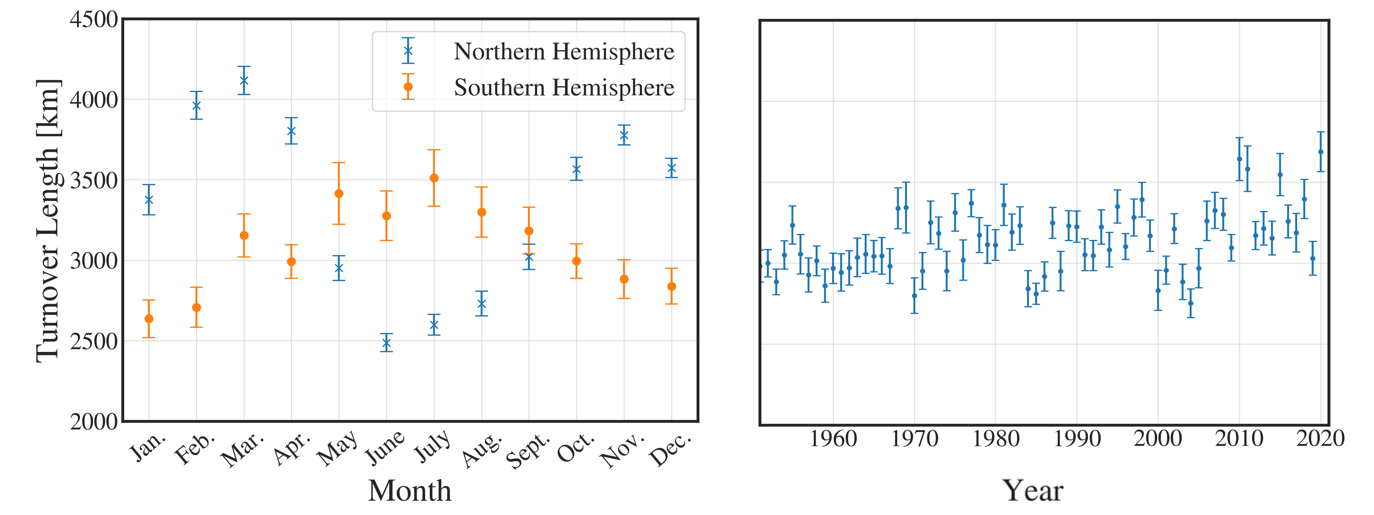}
    \caption{Left: the turnover length scale as a function of months for the northern and southern hemisphere [averaged over the period 1950-2020]. Right: the turnover length scale from 1950-2020 for entire globe. We see the smallest turnover-lengths in summer months, with the larger turnover-length in winter. Furthermore, in recent decades the turnover length has increased.}
    \label{fig:Turnover}
\end{figure*}

\subsection{The Temporal Evolution of Turnover Length}\label{app:turnover}
The emphasis on the coherent temporal evolution of the power within the cascade underlines the growing weather volatility. However, another interesting temporal trend is the changing spatial scale for the transition from flat to cascade. We can determine the intersection between the powerlaw (fitted from modes $10 \leq l \leq 50$) with the average power of modes in the flat regime ($2 \leq l \leq 8$), which yields an estimate of the turnover length. As seen in Fig. \ref{fig:Turnover} this definition proves that the characteristic length scale is dynamically evolving. A noteworthy feature is that the turnover length has increased from $2960 \pm 30$ km (for 1950-1960) to $3230 \pm 60$ km (for 2010-2020). The discrepancy in the length scales for these two periods is more than $3\sigma$ significant.   



\begin{figure*}
    \centering
    \includegraphics[width=0.85\linewidth]{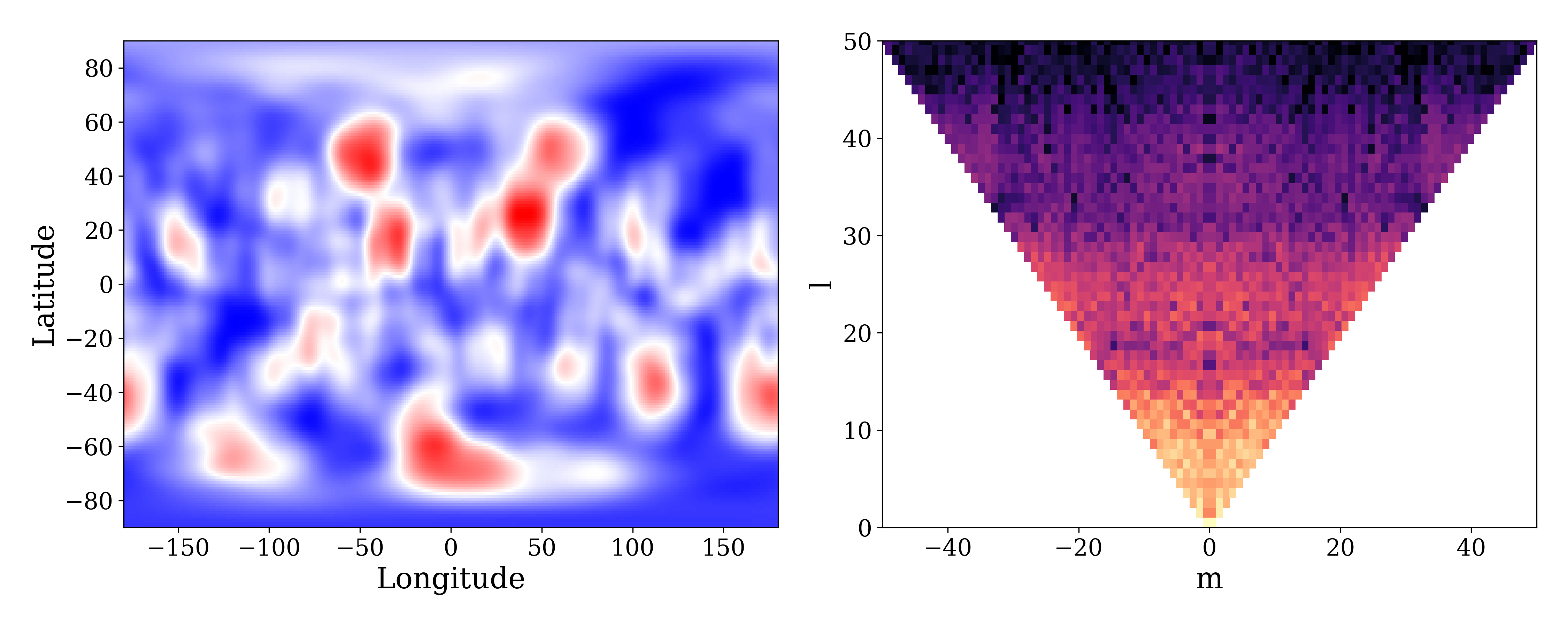}
    \includegraphics[width=0.85\linewidth]{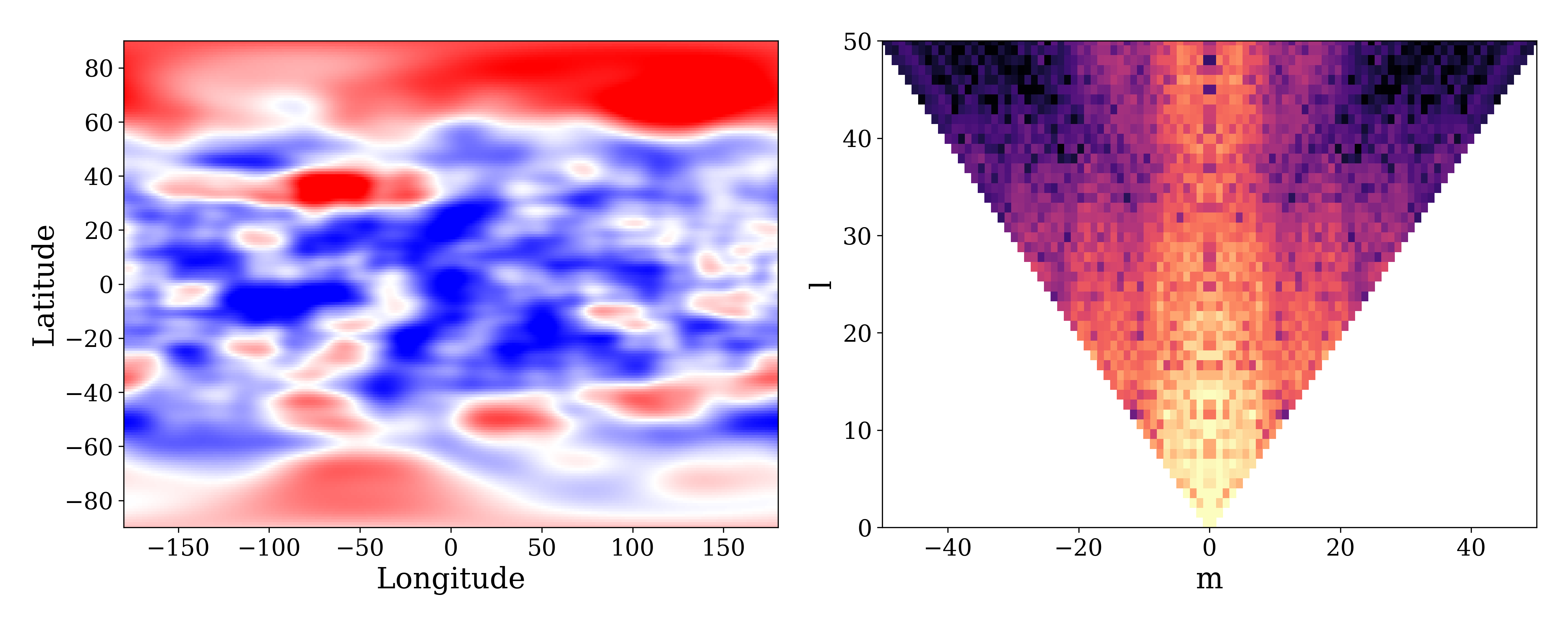}
    \caption{Left panels show simulated temperature field composed from 250 isotropically distributed Gaussian temperature fluctuations with right panels indicating corresponding power spectrum for any spherical harmonic $Y_l^m$. Top illustrates equal gaussian variance along longitudes and latitudes, with bottom showing 5 times larger variance along longitudes than latitudes. Evidently anisotropies of the scale of temperature fluctuations causes an asymmetry in power across $m$.}
    \label{fig:sim_ani}
\end{figure*}

\subsection{Simulated Anisotropic Temperature Fields}\label{app:sim_ani}
To simulate a global temperature field with a prescribed scale of fluctuations, we randomly generate 250 points sampled isotropically across the sphere. Each point represents the center of a Gaussian with the variance setting the scale of fluctuations. Importantly, the variance along latitudes and longitudes can be set independently. The center and spread of the $i$'th Gaussian is denoted respectively ($\phi_i$, $\theta_i$) and ($\sigma_\phi, \sigma_\theta$). Thus, the total temperature field is determined as a superposition of all individual fluctuations: 
\begin{equation}
T(\theta,\phi) = \sum_i{\frac{1}{2 \pi \sqrt{\sigma_\phi \sigma_\theta}} \exp{ \left( -\frac{(\phi-\phi_i)^2 \sin(\theta)^2}{2 \sigma_{\phi}^2} - \frac{(\theta-\theta_i)^2}{2 \sigma_{\theta}^2}\right) }  }
\end{equation}

As seen in Fig. \ref{fig:sim_ani} varying the scale of fluctuations between longitudes and latitudes varies the power across $m$ at fixed $l$. 



\end{document}